\documentclass[12pt]{article}
\usepackage{latexsym,amsmath,amssymb,color}
\usepackage[dvips,colorlinks=true,linkcolor=blue]{hyperref}

\def\F{I\kern-.30em{F}}
\def\P{I\kern-.30em{P}}
\def\E{I\kern-.30em{E}}
\def\build#1_#2^#3{\mathrel{\mathop{\kern
0pt#1}\limits_{#2}^{#3}}}
\def\Sum{\displaystyle\sum}

\def\supp{\mbox{\rm supp}\ }

\newcommand{\R}{\mathbb{R}}

\newcommand{\Z}{\mathbb{Z}}

\newtheorem{theorem}{Theorem}[section]

\newtheorem{corollary}{Corollary}[section]

\newcommand{\Schr}{Schr{\"o}dinger}

\newcommand{\beq}{\begin{equation}}
\newcommand{\eeq}{\end{equation}}
\newcommand{\ba}{\begin{array}}
\newcommand{\ea}{\end{array}}
\newcommand{\bea}{\begin{eqnarray}}
\newcommand{\eea}{\end{eqnarray}}

\begin{document}

\begin{titlepage}

\begin{center}

  {\bf Some New Estimates on the Spectral Shift Function \\
  Associated with Random Schr{\"o}dinger Operators}

  \vspace{0.3 cm}

  \setcounter{footnote}{0}
  \renewcommand{\thefootnote}{\arabic{footnote}}

  {\bf Jean-Michel Combes \footnote{Centre de Physique Th{\'e}orique, CNRS
Marseille, France} }

  \vspace{0.1 cm}

  {D{\'e}partement de Math{\'e}matiques \\
    Universit{\'e} du Sud, Toulon-Var \\
    83130 La Garde, FRANCE}

  \vspace{0.2 cm}

  {\bf Peter D.\ Hislop \footnote{Supported in part by NSF grant
      DMS-0503784.}}

  \vspace{0.1 cm}

  {Department of Mathematics \\
    University of Kentucky \\
    Lexington, KY 40506--0027 USA}

  \vspace{0.2 cm}

  {\bf Fr{\'e}d{\'e}ric Klopp}

  \vspace{0.1 cm}

  {L.A.G.A, Institut Galil{\'e}e\\
    Universit{\'e} Paris-Nord \\
    F-93430 Villetaneuse, FRANCE\\
  et\\
  Institut Universitaire de France}

\end{center}

\vspace{0.2 cm}

\begin{center}
  {\bf Abstract}
\end{center}

\noindent
We prove some new pointwise-in-energy bounds on the
expectations of various spectral shift functions (SSF) associated
with random Schr{\"o}dinger operators in the continuum having Anderson-type
random potentials
in both finite-volume and infinite-volume.
These estimates are a consequence of our new Wegner estimate
for finite-volume random Schr{\"o}dinger operators \cite{[CHK2]}.
For lattice models,
we also obtain a representation of the infinite-volume
density of states in terms of the expectation
of a SSF for a single-site perturbation.
For continuum models,
the corresponding measure, whose density is given by this SSF,
is absolutely continuous with respect to the
density of states
and agrees with it in certain cases.
As an application of one-parameter spectral averaging,
we give a short proof of the classical
pointwise upper bound on the SSF for finite-rank perturbations.

\end{titlepage}




\section{Introduction: The Wegner Estimate and the Spectral Shift
Function}\label{intro}

Some recent analyses of random Schr{\"o}dinger operators have involved three
related
concepts: the Wegner estimate for the finite-volume Hamiltonians,
the spectral shift function (SSF), and the integrated density
of states (IDS). In this note, we prove some new pointwise bounds on the
expectation of some SSFs
that occur in the theory of random Schr{\"o}dinger operators in the 
continuum.
These bounds result from an improved version of the Wegner estimate
\cite{[CHK2]}.
In earlier work \cite{[CHK1],[CHN]}, we used $L^p$-bounds on the SSF in
order
to obtain better estimates on the IDS. In our most recent work, we obtain an
optimal
Wegner estimate directly without using the SSF and found,
as a consequence, new pointwise bounds on the expectation of the SSF.
It has often been conjectured that in the case of ergodic, random,
Schr{\"o}dinger operators of the form
considered here the SSF for a local single-site perturbation should be
in $L_{loc}^\infty (\R)$ once it is averaged
over the random variables on which the disordered potential depends.
We prove this in this note.
We mention that these types of bounds on the SSF also play a motivating role
in the fractional moment method for proving localization in the continuum
\cite{[AENSS]}.
For lattice models, the pointwise bounds on the SSF are a simple consequence
of the
fact that the corresponding perturbations
are finite-rank (cf.\ \cite{[BY],[Yafaev]} and section \ref{maxdis1}).

We first recall a special case of the Wegner estimate proved in
\cite{[CHK2]}
that will be used for the bounds on the SSF.
We refer to \cite{[CHK2]} for the general statement,
valid for arbitrary bounded processes on $\Z^d$,
and the proofs.
The family of \Schr\ operators $H_\omega = H_0 + V_\omega$, on $L^2 (
\R^d)$, is
constructed from a deterministic, $\Z^d$-periodic, background operator $H_0
= (- i \nabla - A_0 )^2 + V_0$.
We consider an
Anderson-type random potential $V_\omega$ constructed from the single-site
potential $u$ as
\begin{equation}\label{anderson1}
V_\omega (x) = \Sum_{j \in \Z^d  } \; \omega_j u ( x - j ) .
\end{equation}
The family of random variables is assumed to be independent, and identically
distributed
(iid).
The results are independent of
the disorder provided it is nonzero.

We define local versions of the Hamiltonians and potentials
associated with bounded regions in $\R^d$.  By $\Lambda_L (x)$, we
mean the open cube of side length $L$ centered at $x \in \R^d$.  For
$\Lambda \subset \R^d$, we denote the lattice points in $\Lambda$ by
${\tilde \Lambda } = \Lambda \cap \Z^d$.
For a cube $\Lambda$, we take $H_0^\Lambda$ and
$H_\Lambda$ (omitting the index $\omega$)
to be the restrictions of $H_0$ and $H_\omega$,
respectively, to the cube $\Lambda$, with periodic boundary
conditions on the boundary $\partial \Lambda$ of $\Lambda$.
We denote by $E_0^\Lambda ( \cdot )$ and $E_\Lambda ( \cdot )$
the spectral families for
$H_0^\Lambda$ and $H_\Lambda$, respectively.  Furthermore, for
$\Lambda \subset \R^d$, let $\chi_\Lambda$ be the characteristic
function for $\Lambda$.  The local potential $V_\Lambda$ is defined by
\beq\label{localpot1}
V_\Lambda (x) = V_\omega ( x) \chi_\Lambda (x) ,
\eeq
and we assume this can be written as
\beq\label{localpot2}
V_\Lambda (x) = \Sum_{j \in {\tilde \Lambda }} \; \omega_j u ( x   - j ).
\eeq
For example, if the support of $u$ is contained in a single unit
cube, the formula (\ref{localpot2}) holds.
We refer to the discussion in \cite{[CHK1]}
when  the support of $u$ is compact, but not necessarily contained
inside one cube. In this case, $V_\Lambda$ can be
written as in (\ref{localpot2}) plus a boundary term of order $| \partial
\Lambda|$
and hence it does not contribute to the large $|\Lambda|$ limit.
Hence, we may assume (\ref{localpot2}) without any loss of generality.
We will also use the local potential obtained from (\ref{localpot2})
by setting all the random variables to one, that is,
\beq\label{localpot3}
{\tilde V}_\Lambda (x) = \Sum_{j \in {\tilde \Lambda }} ~u_j ( x  ),
\eeq
where we will write $u_j(x) = u(x-j)$.

We will always make the following four assumptions:
\vspace{.1in}
\begin{description}
\item[(H1).] The background operator $H_0 = (- i \nabla - A_0)^2 + V_0$ is
a lower semi-bounded,
$\Z^d$-perio\-dic \Schr\ operator with a real-valued, $\Z^d$-periodic,
potential
$V_0$, and a $\Z^d$-periodic vector potential $A_0$. We assume that $V_0$
and
$A_0$ are sufficiently regular so that $H_0$ is essentially
self-adjoint on $C_0^\infty ( \R^d)$.

\item[(H2).] The periodic operator $H_0$  has the unique continuation
property, that is,
for any $E \in \R$ and for any function $\phi \in H^2_{loc} ( \R^d)$, if
$(H_0 - E ) \phi = 0$, and if $\phi$ vanishes on an open set, then $\phi
\equiv 0$.

\item[(H3).] The nonzero, non negative, compactly supported, bounded
  single-site potential $u \in L_0^\infty ( \R^d)$, and it is strictly
  positive on a nonempty open set.

\item[(H4).] The random coupling constants $\{ \omega_j  \;
  | \; j \in \Z^d \}$, are independent and identically distributed.
    The probability distribution $\mu_0$ of $\omega_0$ is
  compactly supported with a bounded density $h_0 \in L_0^\infty ( \R)$.

\end{description}
These imply that the infinite-volume random Schr{\"o}dinger operator
$H_\omega$
is ergodic with respect to the group of $\Z^d$-translations.

Our results also apply to the randomly perturbed Landau Hamiltonian
$H_\omega (\lambda)  =
H_L (B) + \lambda V_\omega$, for $\lambda \neq 0$,
where $V_\omega$ is an Anderson-type potential as in (\ref{anderson1}).
The Landau Hamiltonian $H_L (B)$ on $L^2 ( \R^2)$ is given by
\beq\label{landau1}
H_L (B) = (-i \nabla - A_0 )^2, ~~\mbox{with} ~A_0(x, y) = \frac{B}{2} ( -
y, x).
\eeq
The constant $B \neq 0$ is the magnetic field strength.

Under these assumptions, the Wegner estimate necessary for our purposes
has the following form.

\begin{theorem}\label{wegnermain1}
  We assume that the family of random Schr{\"o}dinger operators
  $H_\omega = H_0 + V_\omega$ on $L^2 ( \R^d)$
  satisfies hypotheses (H1)-(H4).  Then, there exists a locally
  uniform constant $C_W > 0$ such that for any $E_0 \in \R$, and
  $\epsilon \in (0, 1]$, the local Hamiltonians $H_\Lambda$
  satisfy the following Wegner estimate
  \bea\label{wegner11}
  \P \{ \mbox{dist} ( \sigma (H_\Lambda), E_0 ) <
  \epsilon \} & \leq & \E \{ Tr E_\Lambda ( [E_0 - \epsilon, E_0 +
  \epsilon ] ) \}
  \nonumber \\
  &\leq & C_W \epsilon  | \Lambda| .
  \eea
    A similar estimate holds for randomly perturbed Landau Hamiltonians.
\end{theorem}

This theorem immediately implies the Lipschitz continuity of the integrated
density of states \cite{[CHK2]}.
As a consequence, the density of states (DOS) exists and is a locally
bounded function.
In this note, we use Theorem \ref{wegnermain1} to
prove new pointwise bounds on the expectation of the SSF for both
finite-volume
and infinite-volume random Schr{\"o}dinger operators.
We comment on the relation of these results to various results
concerning the SSF for random Schr{\"o}dinger operators in
section \ref{comments}.
We also show that one-parameter spectral averaging can be used to recover
the classical pointwise bound on the SSF for finite-rank perturbations.

\section{Bounds on the Spectral Shift Function for Finite-Volume
Hamiltonians}\label{boundssf1}

We use the result of Theorem \ref{wegnermain1}
to bound the expectation of the
SSF for a single-site perturbation of a finite-volume
Hamiltonian $H_\Lambda$. Since the size of the support of the
perturbation $u$ is of order one relative to $| \Lambda|$,
we expect the SSF to be of order one also.
For a discussion of the relation between the IDS and the SSF, we refer the
reader to \cite{[CHN]} and references therein. A nice review of
results concerning the SSF may be found in \cite{[BY]}.
We recall that for a pair of self-adjoint operators $(H(1), H(0))$,
such that $f(H(1)) - f(H(0))$ is trace-class,
the SSF $\xi (E; H(1), H(0))$ is defined through the trace formula. For
example,
for any
$f \in C_0^1 (\R)$, we have
\beq\label{defn1}
Tr [ f(H(1)) - f(H(0)) ] =  \int_{\R} ~ f'(E) \xi (E; H(1), H(0)) ~dE.
\eeq

We first consider a one-parameter family of self-adjoint operators
$H(\lambda) = H_0 + \lambda V$, with $V \geq 0$, and $\lambda$ uniformly
distributed on $[0, 1]$.
Birman and Solomyak proved a relation (cf.\
\cite{[Simon2]}) between the averaged, weighted, trace of the spectral
projector
$E_\lambda ( \cdot)$ of $H(\lambda)$, and the SSF for the pair $H(1) \equiv
H(\lambda = 1)
$ and $H(0) \equiv H(\lambda = 0) = H_0$.
For any measurable $\Delta \subset \R$, this formula has the form
\beq\label{ssf2}
\int_0^1 ~d \lambda ~Tr ~V^{1/2} E_{\lambda}
(\Delta) V^{1/2} = \int_\Delta ~dE ~\xi (E; H_0 + V , H_0  ) ,
\eeq
whenever all the terms exist. For example, if $V$ is relatively $H_0$-trace
class, then all the terms are well-defined.

We apply this formula as follows. First, we must also make a stronger
hypothesis on the probability distribution than (H4). We will assume:

\begin{description}
\item[(H4').] The random coupling constants $\{ \omega_j  \;
  | \; j \in \Z^d \}$, are independent and identically distributed.
    The probability distribution $\mu_0$ of $\omega_0$ is
  the uniform distribution on $[0, 1]$.
\end{description}

\noindent
As, for some $C>0$, one has $0 \leq \sum_j u_j \leq C$, it follows
trivially that
\beq\label{ssf1}
\sum_{j \in {\tilde \Lambda}} ~Tr u_j^{1/2} E_\Lambda ( \Delta) u_j^{1/2}
\leq C Tr E_\Lambda ( \Delta ).
\eeq

We now consider the effect of the variation of one random variable
$\omega_j$, for $j \in \tilde{\Lambda}$, on the local Hamiltonian.
In formula (\ref{ssf2}), we take $H (0) = H_\Lambda ( \omega_j = 0)$,
$H(1) = H_\Lambda ( \omega_j = 1)$, so that $\lambda = \omega_j$, and $V =
u_j \geq 0$.
We write
$H_{j^\perp}^\Lambda$ for $H_\Lambda$ with $\omega_j = 0$.
The Birman-Solomyak formula (\ref{ssf2}) then has the form
\beq\label{ssf22}
\int_{0}^{1} ~d \omega_j ~Tr ~u_j^{1/2} E_{\Lambda}
(\Delta) u_j^{1/2} = \int_\Delta ~dE ~\xi (E; H_{j^\perp}^\Lambda + u_j,
H_{j^\perp}^\Lambda  ) .
\eeq
Taking the expectation of (\ref{ssf1}) and using formula (\ref{ssf22}),
we obtain
\bea\label{ssf3}
\E \{ \sum_{j \in {\tilde \Lambda}} ~Tr ~u_j^{1/2} E_\Lambda ( \Delta)
u_j^{1/2} \}
& = & \sum_{j \in {\tilde \Lambda}}  \E \left\{ \int_\Delta
~dE ~\xi (E; H_{j^\perp}^\Lambda +  u_j,
H_{j^\perp}^\Lambda  ) \right\} \nonumber \\
& \leq & C \E \{ Tr E_\Lambda ( \Delta ) \} \nonumber \\
& \leq & C_0 ~| \Delta | ~| \Lambda | ,
\eea
where we used the result of the proof of Theorem \ref{wegnermain1} on the
last
line.
We conclude from (\ref{ssf3}) that
\beq\label{ssf4}
\frac{1}{| \Delta |}
\int_\Delta ~dE ~\left\{ \frac{1}{| \Lambda|}
\sum_{j \in {\tilde \Lambda}} \E \{ \xi (E; H_{j^\perp}^\Lambda + u_j,
H_{j^\perp}^\Lambda  ) \} \right\}  \leq C_0 .
\eeq
If the spatially averaged expectation of the SSF is $L_{loc}^1 ( \R)$
in $E$, we can conclude a pointwise bound from (\ref{ssf4}), for Lebesgue
almost every energy $E$, of the form
\beq\label{ssf5}
\E \left\{ \frac{1}{| \Lambda|} \sum_{j \in {\tilde \Lambda}}
\xi (E; H_{j^\perp}^\Lambda + u_j,
H_{j^\perp}^\Lambda ) \right\}  \leq C_0 .
\eeq
In \cite{[CHN]}, we proved that the SSF for local Schr{\"o}dinger operators
with
compactly-supported perturbations is locally-$L^1$, so this pointwise bound
(\ref{ssf5}) holds.
Finally, we observe that due to the periodic boundary conditions on
$\partial \Lambda$
and the $\Z^d$-periodicity of $H_0$, we have that for any $j,k \in
\tilde{\Lambda}$
\beq\label{periodic1}
\E \{ \xi  (E; H_{j^\perp}^\Lambda + u_j,
H_{j^\perp}^\Lambda ) \}  = \E \{ \xi  (E; H_{k^\perp}^\Lambda + u_k,
H_{k^\perp}^\Lambda ) \} ,
\eeq
and consequently it follows from (\ref{ssf5}) that
for any $j \in \tilde{\Lambda}$,
\beq\label{periodic2}
\E \{ \xi (E; H_{j^\perp}^\Lambda + u_j,
H_{j^\perp}^\Lambda ) \}  \leq C_0 .
\eeq

\begin{theorem}\label{expssf1}
Under the hypotheses (H1)-(H4'), the expectation of the
spectral shift
function, corresponding
to the variation of a single site of the
finite-volume Hamiltonian,
is uniformly locally bounded in energy. That is, for any bounded
energy interval, there is a constant $C_I > 0$, independent of $\Lambda$,
so that for Lebesgue almost
every $E \in I$, we have
\beq\label{ssf6}
\E \{ \xi (E; H_{j^\perp}^\Lambda + u_j,
H_{j^\perp}^\Lambda ) \}  \leq C_I ,
\eeq
for any $j \in \tilde{\Lambda}$.
\end{theorem}

In the lattice case, the perturbation $u_j$ is rank-one, so by the general
theory (cf.\ \cite{[Yafaev],[BY]}, or see section \ref{maxdis1}), we have
the bound
\beq\label{ssf7}
\xi (E; H_{j^\perp}^\Lambda + u_j,
H_{j^\perp}^\Lambda )  \leq  1  ,
\eeq
for any $j \in \tilde{\Lambda}$, uniformly in $E \in \R$.


\section{Bounds on the Spectral Shift Function for Infinite-Volume
Hamiltonians}
\label{boundssf2}

We consider the thermodynamic limit of the SSF in (\ref{ssf5}).
The Birkhoff Ergodic Theorem implies that the limit of the right side
of (\ref{ssf5}) is the expectation of the SSF corresponding to the
pair of infinite-volume Hamiltonians
$(H_{0^\perp}, H_{0^\perp} + u_0)$
if we replace $\xi (E; H_{j^\perp}^\Lambda + u_j,
H_{j^\perp}^\Lambda )$ by $\xi (E; H_{j^\perp} + u_j,
H_{j^\perp})$, where $H_{j^\perp}$ is the infinite-volume Hamiltonian with
$\omega_j = 0$.
This is the content of the next proposition.

\begin{theorem}\label{expssf2}
Let $H_{0^\perp}$ be the infinite-volume random Hamiltonian
$H_\omega$ with $\omega_0 = 0$
and assume hypotheses (H1)-(H4').
Then the SSF $\xi( E; H_{0^\perp} + u_0, H_{0^\perp} )$ is well-defined
and $\E \{ \xi( E; H_{0^\perp} + u_0, H_{0^\perp} ) \}  \in L_{loc}^\infty
( \R)$.
\end{theorem}

\noindent
{\bf Proof:}
\noindent
1. We begin with the integrated expression (\ref{ssf4}) and write
\bea\label{tdlssf1}
\lefteqn{ \frac{1}{|\Delta|} \int_\Delta ~ dE ~
\E \left\{ \frac{1}{|\Lambda|}  \sum_{j \in {\tilde \Lambda}}  ~\xi (E;
H_{j^\perp}^\Lambda +  u_j,
H_{j^\perp}^\Lambda  ) \right\} } \nonumber \\
&= & \frac{1}{|\Delta|} \int_\Delta ~ dE ~
\E \left\{ \frac{1}{|\Lambda|} \sum_{j \in {\tilde \Lambda}}  ~\xi (E;
H_{j^\perp} +  u_j,
H_{j^\perp}  ) \right\} + \frac{ \mathcal{E}_\Lambda (\Delta)}{ |\Delta|} ,
\eea
where the error term is
\beq\label{errorssf1}
\mathcal{E}_\Lambda (\Delta) \equiv  \int_\Delta ~ dE ~
\E \left\{ \frac{1}{|\Lambda|} \sum_{j \in {\tilde \Lambda}}  \left( \xi
(E; H_{j^\perp}^\Lambda +  u_j,
H_{j^\perp}^\Lambda  ) - \xi ( E; H_{j^\perp} + u_j , H_{j^\perp} )
\right)  \right\} .
\eeq
We will prove below that $\mathcal{E}_\Lambda (\Delta) \rightarrow 0$ as
$|\Lambda| \rightarrow \infty$.
Assuming this for the moment, it follows from the Birkhoff Ergodic Theorem
and (\ref{tdlssf1}) that
\bea\label{tdlssf2}
\lefteqn{ \lim_{|\Lambda| \rightarrow \infty} \frac{1}{|\Delta|} \int_\Delta
~ dE ~
\E \left\{ \frac{1}{|\Lambda|}  \sum_{j \in {\tilde \Lambda}}  ~\xi (E;
H_{j^\perp} +  u_j,
H_{j^\perp}  ) \right\} } \nonumber \\
&=& \frac{1}{|\Delta|} ~\int_\Delta ~dE ~\E \{ ~\xi (E; H_{0^\perp} +
u_0,
H_{0^\perp} ) \} \leq C_I < \infty.
\eea
In order to justify the interchange of the expectation and the
infinite-volume
limit, we note that the nonnegative series in brackets on the first
line of (\ref{tdlssf2}) converges pointwise a.\ e.\ to the
integrand on the second line of (\ref{tdlssf2}). As the SSF
$\xi(E;H_{j^\perp} +  u_j,
H_{j^\perp}  ) \in L_{loc}^1 (\R)$, the term in the brackets on the right of
the first line
of (\ref{tdlssf2}) is uniformly bounded, so the exchange is justified by the
Lebesgue Dominated Convergence Theorem.
We apply the Lebesgue Differentiation Theorem to the second line of
(\ref{tdlssf2}), since the SSF is in $L^1_{loc} ( \R)$,
and obtain the pointwise bound in Theorem \ref{expssf2}.

\noindent
2. It remains to prove the vanishing of the error term
in (\ref{errorssf1}) in the infinite-volume limit. Using the identity
on the first line of (\ref{ssf3}), we obtain
\beq\label{tdlssf4}
\mathcal{E}_\Lambda ( \Delta) = \E \left\{
\frac{1}{|\Lambda|}
\sum_{j \in {\tilde \Lambda}} \left[ Tr u_j^{1/2} E_\Lambda ( \Delta )
u_j^{1/2} - Tr
u_j^{1/2} E (\Delta) u_j^{1/2} \right] \right\} .
\eeq
We define a local nonnegative measure $\kappa_\Lambda$ by
\beq\label{doslocal1}
\kappa_\Lambda (\Delta) \equiv \frac{1}{|\Lambda|} \E \left\{
\sum_{j \in {\tilde \Lambda}} Tr u_j^{1/2} E_\Lambda ( \Delta ) u_j^{1/2}
\right\} ,
\eeq
and the nonnegative measure $\tilde{\kappa}_\Lambda$,
defined similarly but with the spectral projection $E(\cdot)$
for the infinite-volume Hamiltonian $H_\omega$.
In terms of these local measures,
we can express the right side of (\ref{tdlssf4}) as
\beq\label{tdlssf5}
\mathcal{E}_\Lambda ( \Delta) = \frac{1}{|\Lambda|} [ \kappa_\Lambda
(\Delta)
- \tilde{\kappa}_\Lambda (\Delta) ] .
\eeq
We first prove that the measure $\mathcal{E}_\Lambda ( \cdot)$ converges
vaguely to zero by computing the Laplace transform of the
measure. The Laplace transform $\mathcal{L} ( \mathcal{E}_\Lambda) (t)$ is
easily
seen to be given by
\beq\label{lt1}
\mathcal{L} ( \mathcal{E}_\Lambda) (t) = \frac{1}{|\Lambda|} \E \{ Tr ~
\tilde{V}_\Lambda ( e^{-t
H_\omega}
- e^{-t H_\Lambda } ) \}.
\eeq
Using the Feynman-Kac formula for the heat semigroups, for example, one
easily shows, as in \cite{[Kirsch1]}, that
\beq\label{lt2}
\lim_{|\Lambda| \rightarrow \infty} \mathcal{L} ( \mathcal{E}_\Lambda) (t) =
0,
\eeq
for $t > 0$ pointwise, for a reasonable expanding family of regions
$\Lambda$.
This implies the measure $\mathcal{E}_\Lambda ( \cdot)$ converges vaguely to
zero which, in turn, implies that the right side of (\ref{tdlssf5})
converges to zero. $\Box$

A consequence of this result is an apparently new relationship
between the infinite-volume SSF and the DOS for lattice models. The
analogous relation
for continuum models defines a new measure absolutely continuous with
respect
to Lebesgue measure and to the DOS measure. These results follow easily from
the
proof of Proposition \ref{expssf2}.

\begin{corollary} Let $\nu$ be the DOS measure for the random
Hamiltonian $H_\omega$.
For lattice models, for any Borel set $A \subset \R$, we have
\beq\label{doslattice1}
\nu (A)= \int_A ~dE ~\E \{ \xi (E; H_{0^\perp} + u_0, H_{0^\perp} ) \} .
\eeq
For continuum models, there is a nonnegative measure $\kappa$, absolutely
continuous with respect
to the DOS measure and Lebesgue measure, with distribution given in
(\ref{distribution1}),
so that
\beq\label{doscontinuum1}
\kappa (A)= \int_A ~dE ~\E \{ \xi (E; H_{0^\perp} + u_0, H_{0^\perp} ) \} .
\eeq
For any closed bounded interval $I \subset \R$, there are constants $0 < c_I
\leq C_I < \infty$, so that
for any Borel set $A \subset I$, we have
\beq\label{equiv1}
0 \leq \kappa (A) \leq c_I | A |, ~\mbox{and} ~ 0 \leq \kappa (A) \leq C_I
\nu(A).
\eeq
\end{corollary}

\noindent
{\bf Proof:}
>From the Birkhoff Ergodic Theorem, and expression (\ref{ssf22}),
we can express the integral on the right in (\ref{tdlssf2})
in terms of a positive measure $\kappa$ as follows
\bea\label{tdlssf3}
\int_\Delta ~dE ~\E \{ ~\xi (E; H_{0^\perp} +  u_0,
H_{0^\perp} ) \} &=&
\E \left( \lim_{|\Lambda| \rightarrow \infty}
\frac{1}{|\Lambda|} ~\sum_{j \in {\tilde \Lambda}} Tr u_j^{1/2}
E ( \Delta ) u_j^{1/2} \right)
  \nonumber \\
&=& \E \{ Tr u_0^{1/2} E( \Delta ) u_0^{1/2} \} \nonumber \\
&\equiv &  \kappa ( \Delta) ,
\eea
where $\kappa ( \cdot)$ is the nonnegative measure with distribution
function given by
\beq\label{distribution1}
K(E) \equiv \E \{ Tr u_0^{1/2} P(E) u_0^{1/2} \} ,
\eeq
where $P(E)$ is the spectral family for $H_\omega$.
For the lattice case,
this measure is just the DOS measure, since $u_0 = \delta_0$,
so that $\E \{ ~\xi (E; H_{0^\perp} +  u_0,
H_{0^\perp} ) \}$ is a representation of the DOS.
It follows immediately from (\ref{tdlssf3}) and Theorem \ref{expssf2} that
for any closed bounded interval $I \subset \R$, there exists a finite
constant
$0 < C_I < \infty$, so that for any
Lebesgue measurable set $A \subset I$, we have
\beq\label{ssfdensity1}
0 \leq \kappa ( A) = \int_A  ~dE ~\E \{ ~\xi (E; H_{0^\perp} +  u_0,
H_{0^\perp} ) \} \leq C_I |A|.
\eeq
Lebesgue measure,
It remains to prove that $\kappa$ is bounded above by the DOS measure. This
implies
the absolute continuity with respect to $\nu$. We simply note that there
exists a constant $0 < C_0 < \infty$,
depending only on $u$, so that
\beq\label{upperbound1}
0 \leq \sum_{j \in \tilde{\Lambda}} Tr u_j^{1/2} E_\Lambda ( \Delta )
u_j^{1/2} \leq
C_0 ~Tr E_\Lambda ( \Delta ) ,
\eeq
and recall the definition of the DOS measure. This implies that
$0 \leq \kappa (A) \leq  C_I \nu (A)$, for $A \subset I \subset \R$.
$\Box$.

This measure $\kappa$ is similar to the DOS measure for continuum models.
The distribution function for the DOS for continuum models is given by
$N(E) = \E \{ Tr \chi_{\Lambda_1(0)} P(E) \chi_{\Lambda_1(0)} \}$.
The measure $\kappa$ is equivalent to the DOS measure $\nu$
if the single-site potential satisfies $c_0 \chi_{\Lambda_1 (0)} \leq u$,
for some $c_0 > 0$, and it is equal to $\nu$ in the special case that $u =
\chi_{\Lambda_1 (0)}$.
The equivalence of measures means that there are constants $C_0 , c_0 > 0$
so that
\beq\label{doscompare1}
c_0 \nu(A) \leq \kappa (A) \leq C_0 \nu (A),
\eeq
for all Borel subsets $A \subset \R$.


\section{Comments}\label{comments}

We make three comments on various other results concerning the SSF
associated with
random Schr{\"o}dinger operators that have recently
occurred in the literature related to random Schr{\"o}dinger operators.
For deterministic Schr{\"o}dinger operators, {\it pointwise bounds} are 
known
only in a few specific
cases, such as finite-rank perturbations (cf.\ \cite{[Yafaev],[BY]} and
Theorem \ref{ssf1rankN} below)
or perturbations of the Laplacian on $L^2 ( \R^d)$ by sufficiently smooth
potentials \cite{[Sobolev1]}.

\subsection{Related Results on the Averaged SSF}

Bounds on the $L^p$-norm of the SSF, for $0 < p \leq 1$,
were proved in \cite{[CHN]} and improved in
\cite{[HS1]}.
More recently,
Hundertmark, et.\ al.\ \cite{[HKNSV]}, obtained some new integral bounds
on the SSF that indicate that one cannot expect that,
in general, the SSF is locally bounded.
Indeed, Kirsch \cite{[Kirsch3],[Kirsch4]}
proved that if the Dirichlet Laplacian
in $\Lambda_L$, a cube
of side length $L$ centered at the origin, is perturbed by a
nonnegative, bounded potential supported inside the unit cube $\Lambda_1$,
then the corresponding
finite-volume SSF, at any positive energy,
diverges as $L \rightarrow \infty$.
Raikov and
Warzel \cite{[RW1]} considered the SSF for the
Landau Hamiltonian (\ref{landau1})
and a perturbation by a compactly-supported potential. They showed
that the SSF diverges at the Landau energies.

The averaged SSF is expected to be better behaved. In addition to the
pointwise bounds of Theorems \ref{expssf1}
and \ref{expssf2}, Aizenman, et.\ al. \cite{[AENSS]} proved
an interesting bound on a spectral shift function related to the
ones treated here. They consider the SSF $\xi(t, E)$ for a pair of
Hamiltonians $H_t = H_0  + tV$ and
$H_t + U$, where $V$ and $U$ are nonnegative bounded potentials such that
$V$ is strictly positive
on a neighborhood of the support of $U$. Specifically, for any $\delta > 0$,
we define the set $Q_\delta \equiv \{ x \in \R^d ~|~ \mbox{dist} ( x , \supp
(U)) < \delta \}$.
We then require that $V$ be strictly positive on $Q_\delta$.

\begin{theorem}\label{aenss1}
For any $0 < s < \mbox{min} ( 2/d, 1/2)$, there is a finite positive
constant $C_{s, \delta} > 0$
so that the SSF $\xi (t, E)$ satisfies the bound
\beq\label{ssfbd01}
\int_0^1 ~| \xi(t, E) |^s ~dt \leq C_{s, \delta} \|U \|_\infty ( 1 + |E -
E_0| +
\|V\|_\infty )^{2s(d+1)} ,
\eeq
where $E \geq E_0 \equiv \inf \sigma (H_0)$.
\end{theorem}

\subsection{Spectral Shift Density}

Kostrykin and Schrader \cite{[KS1],[KS2]} introduced the {\it spectral
shift density} (SSD) that is closely related to the integrated density of
states.
The SSD is the density of a measure $\Xi$ obtained by the thermodynamic
limit
\beq\label{ssd1}
\int_{\R} ~g(\lambda) d \Xi (\lambda) = \lim_{| \Lambda | \rightarrow
\infty}
\int_{\R} ~g(\lambda) \frac{ \xi ( E ; H_0 + \chi_{\Lambda}
V_\omega ) }{ | \Lambda | } .
\eeq
Note that the size of the perturbation $\chi_\Lambda V_\omega$
is of order $|\Lambda|$. They prove that the SSD $\tilde{\xi} (E)$ is given
as
\beq\label{ssd2}
\tilde{\xi}(E) = N_0 ( E) - N(E) , ~\mbox{a.\ e.} ~E \in \R,
\eeq
where $N_0(E)$ is the IDS of $H_0$ and $N(E)$ is the IDS of $H_\omega$.

\subsection{A Pointwise Bound on the SSF for Finite Rank Perturbations}
\label{maxdis1}

We consider the SSF for a finite-rank perturbation.
Let $B \geq 0$
be a nonnegative finite-rank operator with rank $N$. Let $H_s = H_0 + s B$
be the one-parameter perturbation of a self-adjoint, lower-semibounded
operator $H_0$.
The variable $s \in [0,1]$ is uniformly distributed.
We consider the SSF $\xi ( E ; H_1, H_0)$ and recover the classical
pointwise upper
bound $N$ usually obtained by other methods
cf.\ \cite{[BY],[Yafaev]}.

\begin{theorem}
\label{ssf1rankN}
The spectral shift function for the pair of self-adjoint operators
$(H_1, H_0)$, where $0 \leq B \equiv H_1 - H_0$ is a finite-rank
operator of rank $N < \infty$, satisfies the bound
\beq\label{ssf1est1} 0 \leq \xi (E; H_1, H_0) \leq N .  \eeq
\end{theorem}

\noindent
{\bf Proof:}
Let $f \in C^1_0 ( \R)$ and consider the formula for the SSF:
\bea\label{trace1}
Tr ( f(H_1 ) - f(H_0) ) &=& - \int_\R f'(E) \xi (E; H_1 , H_0 ) ~dE
\nonumber
\\
&=& \int_0^1 \frac{d}{ds} Tr f(H_s) ~ds \nonumber \\
&=& \int_0^1 ~ds ~Tr B^{1/2} f' (H_s) B^{1/2} \nonumber \\
&=& \sum_{j = 1}^N \int_0^1 ~ds ~\langle \phi_j , B^{1/2} f'(H_s) B^{1/2}
\phi_j \rangle
\eea
Let $E_s ( \cdot )$ be the spectral family for $H_s$. The matrix
element in (\ref{trace1}) is written as
\bea\label{trace2}
\langle \phi_j , B^{1/2} f'(H_s) B^{1/2}
\phi_j \rangle &= & \int_\R ~f'(\lambda) ~d \langle B^{1/2} \phi_j , E_s
(\lambda) B^{1/2} \phi_j \rangle \nonumber \\
& = & \int_\R ~f'(\lambda) ~d\mu_{H_s}^{\psi_j} (\lambda),
\eea
where $\psi_j \equiv B^{1/2} \phi_j$ and $\mu_{H_s}^{\psi_j}$ is the
corresponding spectral measure for $H_s$ and $\psi_j$.
We divide the support of $f'$ into $p$ subintervals $\Delta_k$ and bound
the absolute value of the integral over $\lambda$ in (\ref{trace2}) from
above as
\beq\label{approx1}
\left| \int_{\R} ~f' (\lambda) ~d \mu_{H_s}^{\psi_j} (\lambda) \right|
\leq \sum_{k=1}^p  ~|f' (x_k )| ~\mu_{H_s}^{\psi_j} ( \Delta_k) ,
\eeq
where $x_k \in \Delta_k$ is such that $|f(x_k)| = \sup_{x \in \Delta_k} |f'
(x )|$.
Inserting this into the integral over $s$ in (\ref{trace1}), we see that it
remains to
estimate
\beq\label{trace3}
\int_0^1 ~\mu_{H_s}^{\psi_j} (\Delta_k ) ~ds = \int_0^1 ~\langle B^{1/2}
\phi_j , E_s (\Delta_k) B^{1/2} \phi_j \rangle
~ds.
\eeq
For bounded probability distributions with compact support, the integral on
the
right in (\ref{trace3}) was estimated in \cite{[CH1]}. The result is
\beq\label{trace33}
\int_0^1 ~\langle B^{1/2} \phi_j , E_s (\Delta_k) B^{1/2} \phi_j \rangle
~ds \leq | \Delta_k|,
\eeq
since $\| \psi_j \| \leq 1$.
Combining this bound with (\ref{trace1})--(\ref{trace2}) and recalling the
approximation (\ref{approx1}), we obtain
\bea\label{trace7}
\left| \int_\R f'(E) \xi (E; H_1 , H_0 ) ~dE \right| & \leq &
\sum_{j=1}^N ~\sum_{k=1}^p ~|f' (x_k )| ~| \Delta_k| \nonumber \\
& \leq &  N \| f' \|_1 ,
\eea
which, extending the estimate to any $f \in L^1 ( \R)$, we conclude that
\beq\label{trace8}
| \xi ( E; H_1, H_0 ) | \leq  N ,
\eeq
proving the result. $\Box$


\end{document}